\documentclass[conference]{IEEEtran}
\IEEEoverridecommandlockouts
\usepackage{cite}
\usepackage{textcomp} 
\usepackage{amsmath,amssymb,amsfonts}
\usepackage{algorithmic}
\usepackage{graphicx}
\usepackage{xcolor}
\def\BibTeX{{\rm B\kern-.05em{\sc i\kern-.025em b}\kern-.08em
    T\kern-.1667em\lower.7ex\hbox{E}\kern-.125emX}}
\begin{document}

\title{Dynamic Incentive Strategies for Smart EV Charging Stations: An LLM-Driven User Digital Twin Approach}

\author{\IEEEauthorblockN{Yichen Sun}
\IEEEauthorblockA{\textit{Shanghai University of Electric Power} \\
Shanghai, China \\
yichensun@mail.shiep.edu.cn}
\and
\IEEEauthorblockN{Chenggang Cui}
\IEEEauthorblockA{\textit{Shanghai University of Electric Power} \\
Shanghai, China \\
cgcui@shiep.edu.cn}
\and
\IEEEauthorblockN{Chuanlin Zhang}
\IEEEauthorblockA{\textit{Shanghai University of Electric Power} \\
Shanghai, China \\
clzhang@shiep.edu.cn}
\and
\IEEEauthorblockN{Chunyang Gong}
\IEEEauthorblockA{\textit{Shanghai University of Electric Power} \\
Shanghai, China \\
gongchunyang@shiep.edu.cn}
}

\maketitle

\begin{abstract}
This paper presents an enhanced electric vehicle demand response system based on large language models, aimed at optimizing the application of vehicle-to-grid technology. By leveraging an large language models-driven multi-agent framework to construct user digital twins integrated with multidimensional user profile features, it enables deep simulation and precise prediction of users' charging and discharging decision-making patterns. Additionally, a data- and knowledge-driven dynamic incentive mechanism is proposed, combining a distributed optimization model under network constraints to optimize the grid-user interaction while ensuring both economic viability and security. Simulation results demonstrate that the approach significantly improves load peak-valley regulation and charging/discharging strategies. Experimental validation highlights the system's substantial advantages in load balancing, user satisfaction and grid stability, providing decision-makers with a scalable V2G management tool that promotes the sustainable, synergistic development of vehicle-grid integration.

\end{abstract}


\section{Introduction}

Electric vehicles (EVs) serve as essential elements in demand response (DR) systems, acting as flexible distributed energy resources that facilitate clean energy transition and emission reduction \cite{b1}. The integration of EVs with the energy sector via vehicle-to-grid (V2G) technology is emphasized by \cite{b2} for its potential to enhance wind energy utilization and decrease CO2 emissions.

With the rapid proliferation of Electric Vehicles (EVs), grid stability and energy distribution confront unprecedented challenges, which has propelled the development of advanced charging infrastructures and intelligent management strategies. Researchers have proposed decentralized coordination mechanisms utilizing smart algorithms and game theory to optimize Vehicle-to-Grid (V2G) applications. In \cite{b3}, a method for providing V2G regulation services through distributed coordination of electric vehicles is explored, indicating that intelligently scheduling charging times can balance load fluctuations, improve voltage profiles, and reduce network losses without necessitating additional investments. Reference \cite{b4} further advances this by proposing optimal schemes for electric vehicle charging and discharging schedules using metaheuristic algorithms, emphasizing the significance of V2G approaches in reducing costs and supporting the grid. Concurrently, \cite{b5} investigates decentralized vehicle-grid control strategies for primary frequency regulation considering charging demands, demonstrating how these strategies enhance the grid's adaptability to renewable energy fluctuations.

EV users are pivotal in the successful deployment of V2G technology. Research focused on optimizing V2G integration through demand response employs real-time pricing and considers user preferences, using incentives or price signals to encourage participation \cite{b6},\cite{b7}. This aims to balance grid loads, reduce peak demand, and improve societal welfare and grid efficiency. However, these studies often overlook individual behavioral diversity. Recent research highlights that psychological and economic factors, such as financial status, risk preference, and price elasticity, significantly influence user participation decisions \cite{b8},\cite{b9}. By applying psychology and behavioral economics frameworks, In \cite{b10},researchers have revealed differences among user groups in demand response scenarios, though this approach adds complexity and may limit broader applicability due to the nuanced differences among users .

In this paper, an enhanced Electric vehicles demand response system has been proposed, aiming to promote orderly charging and the application of V2G technology. This system employs a multi-agent framework driven by a large language model (LLM), which comprehensively considers users' basic information, psychological, and economic factors to generate precise user profiles. By integrating real-time data from the environment and electric vehicle charging station (EVCS), the system dynamically adjusts strategies and conducts an in-depth analysis of EV user behavior to predict users' willingness to participate in DR. 

The system, through the analysis of user behavior and psychological drivers, achieves a profound understanding of personalized needs, thereby overcoming the limitations of traditional DR methods that neglect user heterogeneity. By employing dynamically adjusted economic incentives, psychological barriers to user participation in V2G programs are effectively addressed, balancing user engagement with load fluctuation control while significantly reducing the high costs associated with V2G pilot projects. The main contributions of this paper are as follows:
\begin{enumerate}
\item User Behavior Simulation: A multi-agent system driven by LLMs is proposed to analyze EV user behavior. The system features a User Profile Generation Agent, consisting of an Information Analysis Agent and a User Psychological Analysis Agent. The former constructs a basic profile from personal data, while the latter applies psychological and economic theories to infer preferences and behavioral patterns, producing a comprehensive profile. A Decision-Making Agent then integrates these profiles with real-time data on EVCS status, environmental conditions, and incentive policies to derive optimal strategies. This collaborative framework enables refined behavioral modeling and precise decision-making analysis.

\item Dynamic Incentive Mechanism: A DR system enhanced by LLMs is proposed, which not only considers users' basic information but also deeply analyzes psychological traits and economic conditions to generate precise behavioral decisions and simulate user behavior in depth. Through the dynamic adjustment of economic incentive mechanisms, the system effectively promotes the orderly charging of EVs and participation in V2G services, optimizing the grid load distribution while avoiding the emergence of new peak loads caused by excessive incentives.
\end{enumerate}

\section{SYSTEM DESCRIPTION AND PROBLEM STATEMENT}

\subsection{Electric Vehicle Charging Station Model}

The study introduces an innovative model for an EVCS that incorporates both charging and discharging services. The EVCS features multiple charging points, and EVs actively communicate their current battery state (\(SOC_n^c\)), target battery state (\(SOC_n^t\)), and their distance from the station prior to arrival. This information allows the EVCS to predict charging demand and estimate the required charging time (\(\tau_n^c\)) accordingly.

The charging energy demand for each vehicle, \(E_n^c\), is calculated based on the difference between the target and current battery states, as well as the vehicle’s battery capacity, as follow:

{\footnotesize
\begin{equation}
\begin{aligned}
 E_n^c = \left( SOC_n^t - SOC_n^c \right) \cdot C_n^{\text{battery}}
\end{aligned}\end{equation}}Subsequently, the charging duration, \(\tau_n^c\), is determined by the following formula:

{\footnotesize
\begin{equation}
\begin{aligned}\tau_n^c = \left\lceil \frac{ E_n^c}{\eta_n^c \Delta P_0^c } \right\rceil
\end{aligned}\end{equation}}\(\eta_n^c\) represents the vehicle's charging efficiency, and \(\Delta P_0^c\) is the nominal charging power of the station's charging point. \(E_n^c\) denotes the charging energy demand for each vehicle.

To manage peak load periods, the EVCS uses historical data to forecast demand surges and issues discharge requests to users, offering V2G incentives, \( R_{\text{V2G}} \), for specific time windows. Users can opt into DR programs based on their energy needs. For each kWh discharged to the grid, users are compensated at a rate of \( r_{\text{V2G}} \) per kWh, motivating participation in the V2G system.
\begin{figure}[htbp]
\includegraphics[width=\columnwidth]{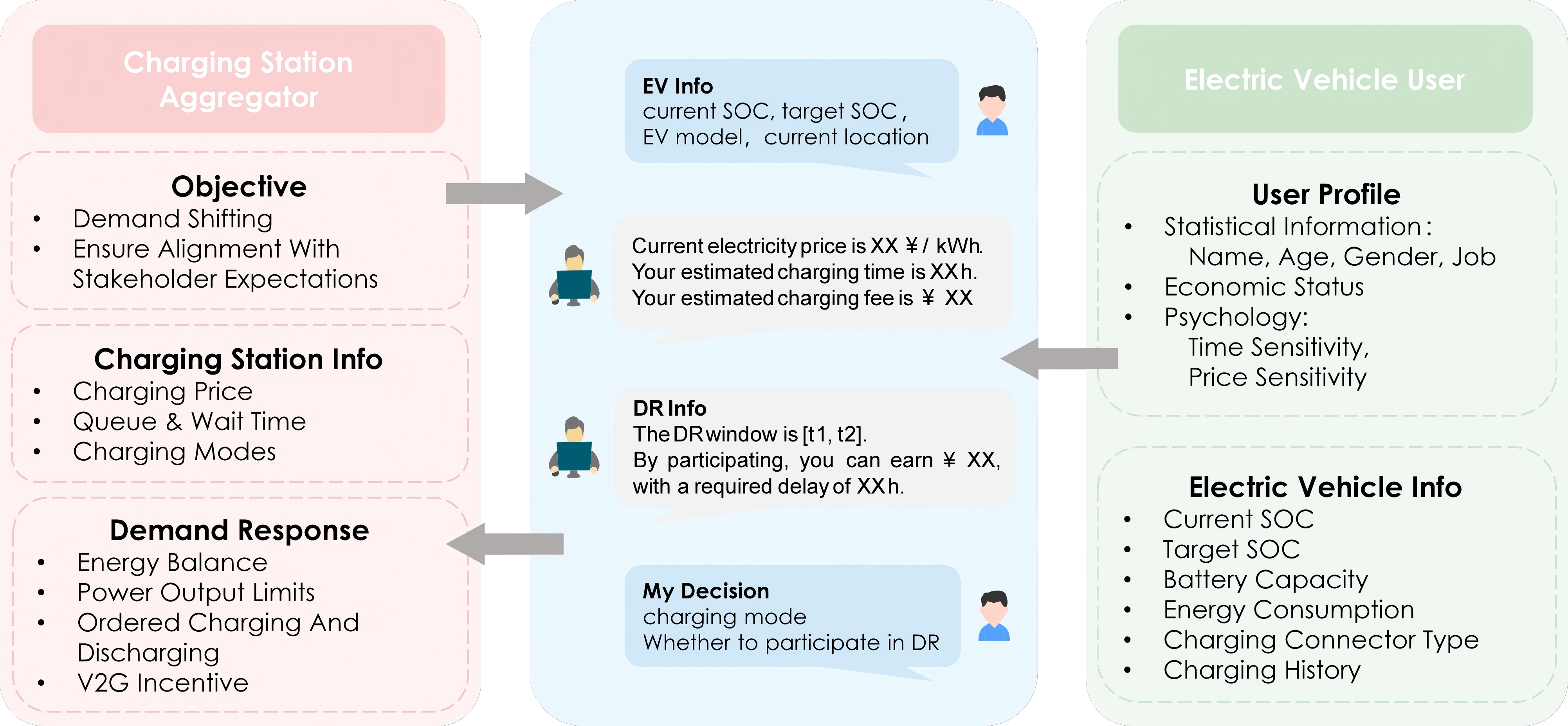}
\caption{Electric vehicle charging station model.}
\label{fig:Electric vehicle charging station model.}
\end{figure}

The current time is denoted as \( t_{\text{current}} \), with peak start and end times represented as \( t_{\text{peak}}^{\text{start}} \) and \( t_{\text{peak}}^{\text{end}} \), respectively. The EVCS offers two charging modes, and users can choose based on their needs. If current time exceeds peak end time, the user can no longer participate in DR. If current time falls within the peak window, users can opt to participate in DR. 

Before peak start time, users wishing to engage in DR must delay the charging start time to within the peak window to benefit from incentives.

The delay time \( t_{\text{delay}} \), caused by participation in DR, is calculated as follows for discharging:
{\footnotesize
\begin{equation}
\begin{aligned}t_{\text{delay}} = 
\begin{cases} 
t_{\text{discharge}}, & \text{if } t_{\text{current}} \in [t_{\text{peak}}^{\text{start}}, t_{\text{peak}}^{\text{end}}] \\
t_{\text{discharge}} + (t_{\text{current}} - t_{\text{peak}}^{\text{start}}), & \text{if } t_{\text{current}} < t_{\text{peak}}^{\text{start}} \\
0, & \text{if } t_{\text{current}} > t_{\text{peak}}^{\text{end}}
\end{cases}
\end{aligned}\end{equation}}Here, \( t_{\text{discharge}} \) is the discharge duration, calculated by:
{\footnotesize
\begin{equation}
\begin{aligned}
t_{\text{discharge}} = \frac{SOC_c \cdot C_{\text{battery}} - (D \cdot \varphi)}{\eta_n^d \Delta P_0^d}
\end{aligned}\end{equation}}Where, \( SOC_c \) is the current state of charge of the battery.
\( C_{\text{battery}} \) is the battery capacity.
\( D \) is the distance from the current location to the EVCS (in kilometers).
\( \varphi \) is the energy consumption per 100 km.
\(\eta_n^d\) represents the vehicle's discharging efficiency,
\(\Delta P_0^d\) is the nominal discharging power of the station's charging point.

\subsection{Demand Response Model}

The DR model adjusts energy demand to balance supply and demand. It uses time-of-use pricing and incentives to encourage users to optimize charging, with users earning rewards for V2G during peak hours and charging during off-peak hours. The model factors in user behavior, vehicle status, environmental data, and grid load to maximize station revenue and ensure load balance

The objective function \( \mathcal{J}_{\text{EVCS}} \) aims to maximize the net profit of the EVCS. Revenue comes from three sources: V2G rewards \( r_{\text{V2G}} \) during the V2G window (e.g., 17:00-22:00 pm), the charging costs to meet vehicle demand, and the cost of purchasing electricity from the grid.

{\footnotesize
\begin{equation}
\begin{aligned}\mathcal{J}_{\text{EVCS}} = \sum_{n=1}^{N} \sum_{t \in [T_0, T_1]} \left( r_{\text{V2G}} \cdot \Delta E_{n}(t) \right) \\- \sum_{n=1}^{N} \int_0^{T_{\text{end}}} P_{n}^c(t) \cdot p^{\text{c}}_{\text{n,t}} \, dt \\+ \sum_{n=1}^{N} \int_0^{T_{\text{end}}} P_{n}^c(t) \cdot p^{\text{grid}}_{\text{t}} \, dt 
\end{aligned}\end{equation}}where, the first part refers to the V2G revenue, where the EVCS earns rewards by feeding electricity from EV batteries back into the grid within the time window \([T_0, T_1]\). The reward amount \(r_{\text{V2G}}\) represents the payment per kWh of electricity fed back to the grid. At time \(t\), the amount of energy discharged by the participating EV is denoted as \(\Delta E_{n}(t)\) .The second part is the charging cost, calculated based on the EV's charging power and the time-varying charging price \(p_{\text{charge}}^{\text{n,t}}\). \(P_n^c(t)\) represents the charging power at time \(t\),which is equal to the \(\eta_n^c\Delta P_0\) ,and \(p_{\text{n,t}}^{\text{c}}\) is the charging price for the \(n\)-th EV at time \(t\) .
    The third part is the electricity procurement cost, calculated based on the grid's time-of-use electricity price \(p^{\text{grid}}_{\text{t}}\).

The objective function \( \mathcal{J}_{\text{user}} \) is designed to maximize the user's profit by optimizing the battery's charging and discharging operations to achieve the highest economic return. It incorporates the user's revenue, charging costs, the cost of deferring charging time, and the discharge-related efficiency loss.

{\footnotesize
\begin{equation}
\begin{aligned}\mathcal{J}_{\text{user}}= \int_{T_0}^{T_1} P(t) \cdot R_{\text{V2G}} \, dt - \int_0^{T_{\text{end}}} P_n(t) \cdot p_{\text{charge}} \, dt\\ - C_{\text{charge}} \cdot T_{\text{delay}} - \int_{T_0}^{T_1} P(t) \cdot \left( 1 - \frac{1}{\eta_{\text{d}}} \right) \cdot  C_{\text{loss}} \, dt 
\end{aligned}\end{equation}}In the objective function \(\mathcal{J}_{\text{EVCS}}\), V2G revenue and charging costs have been elaborated in the first two components. The third component introduces the cost of delayed charging, where \(C_{\text{charge}}\) denotes the cost per unit time of delayed charging, and \(T_{\text{delay}}\) represents the duration of the charging delay.

The loss due to battery discharge inefficiency is accounted for by the factor \(1 - \frac{1}{\eta_{\text{discharge}}}\), reflecting the proportion of energy loss due to suboptimal discharge efficiency \(\eta_{\text{discharge}}\). Additionally, the cost of energy loss per unit power per unit time is denoted by \(C_{\text{loss}}\), assessing the economic impact on users from reduced discharge efficiency.

{\footnotesize
\begin{equation}
\begin{aligned}
E_{\text{current}} + \int_0^{T_{\text{end}}} \Delta E(t) \, dt = E_{\text{target}} 
\end{aligned}\end{equation}}
{\footnotesize
\begin{equation}
\begin{aligned}
E_{\text{current}} + \int_0^{T_{\text{end}}} \Delta E(t) \, dt \leq E_{\text{max}} 
\end{aligned}\end{equation}}
{\footnotesize
\begin{equation}
\begin{aligned}
E_{\text{current}} + \int_0^{T_{\text{end}}} \Delta E(t) \, dt \geq E_{\text{min}}  
\end{aligned}\end{equation}}

{\footnotesize
\begin{equation}
\begin{aligned}
|P(t)|\leq P_{\text{max}} \quad \forall t \in [0, T_{\text{end}}]  
\end{aligned}\end{equation}}

{\footnotesize
\begin{equation}
\begin{aligned}
\left\{ \begin{aligned} P(t) & \geq 0 \quad \text{or} \quad P(t) \leq 0, \quad t \in [T_0,T_1] \\ P(t) & \geq 0, \quad \forall t \notin [T_0,T_1] \end{aligned} \right. 
\end{aligned}\end{equation}}

In the DR model of EVCS. Energy balance constraints (7),(8) and (9) ensure the final energy state meets user needs; maximum and minimum energy bounds guarantee battery safety; power output limits (10) prevent equipment overload and maintain grid stability; constraint (11) on charging/discharging behavior specifies power direction within certain periods and ensures chargers do not charge and discharge simultaneously at any point in time.

\begin{figure}[htbp]
\includegraphics[width=\columnwidth]{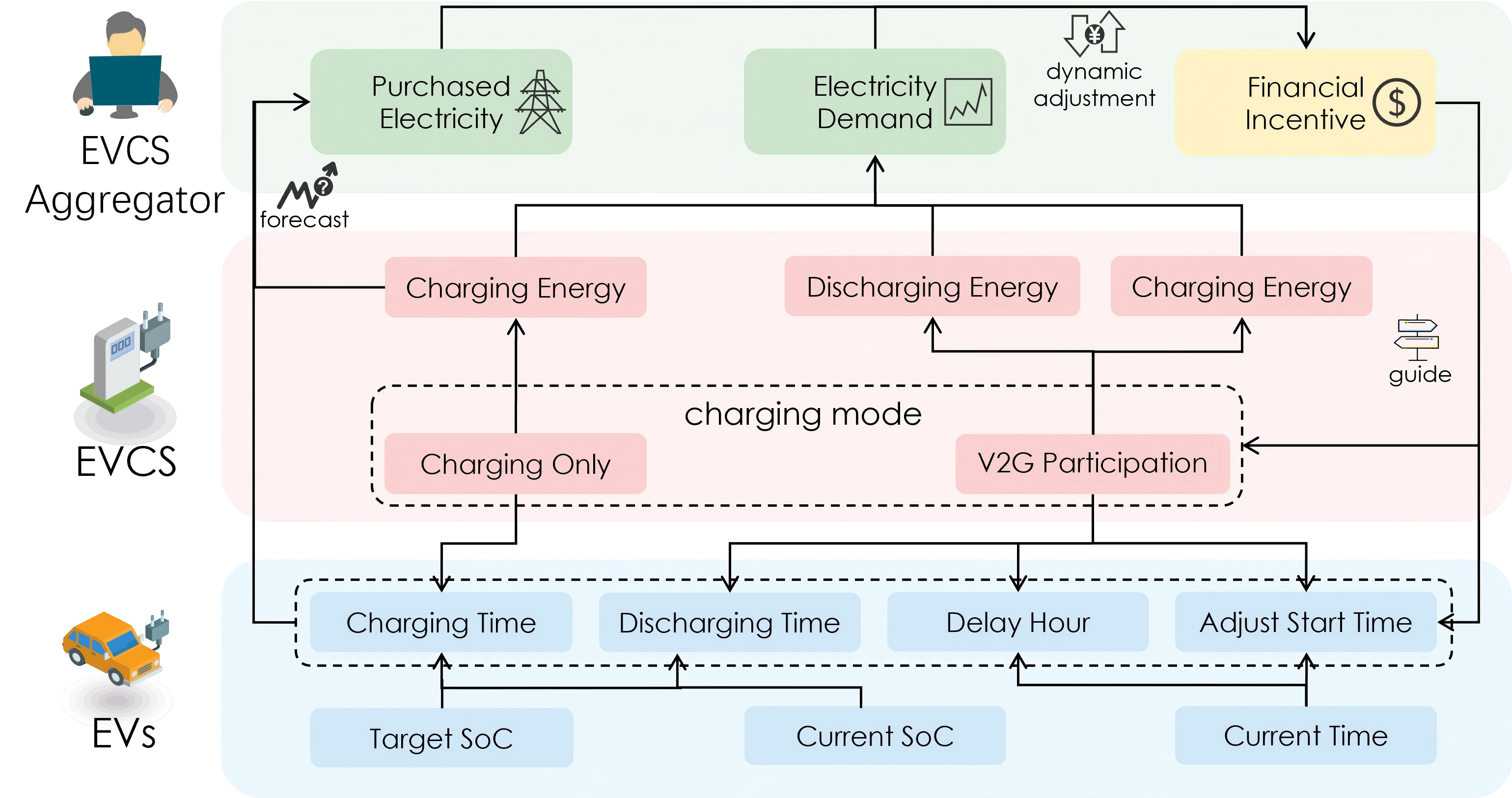}
\caption{Demand response model.}
\label{fig:Demand response model.}
\end{figure}

\section{LLM-Driven Digital Twin of EV Users for Enhanced Behavior Modeling}

In DR research and implementation, user behavior is a critical factor. With the rapid adoption of EVs, both governments and academia have recognized their potential as flexible, controllable distributed energy resources, driving the development of V2G technology. However, the widespread adoption of V2G faces multiple challenges, including technical bottlenecks, unclear business models, high initial investment, insufficient policy support, and low public awareness. Additionally, the diversity and uncertainty of user behavior exacerbate these issues. To address these challenges, this paper introduces a data-driven digital twin of EV users powered by LLMs, which simulates real user behavior and offers new insights and solutions for overcoming these barriers.

\subsection{User Profile Generation Agent}\label{AA}

User profiling is a critical step in the personalized decision-making process for EV users' digital twins. Unlike traditional user modeling methods that rely on generalized approaches, this process enables individualized behavior simulation, offering more accurate and differentiated demand forecasting and decision support. 

Three LLM-powered intelligent agents are developed: a user information agent, an EV information agent, and a psychological trait agent.Through the coupled interaction of these agents, user data is progressively analyzed, leading to the generation of detailed User Profiles and personalized needs. This approach not only considers basic user information but also integrates multidimensional data, such as psychological traits and economic status, to construct a dynamic, highly precise user profile, thereby facilitating personalized decision-making.

\begin{itemize}
\item The Basic Info Agent gathers and analyzes users' demographic data (such as age, gender, and economic status) to construct an initial profile. By classifying users based on occupation and income, it segments them into distinct economic tiers. Using this data, the agent predicts charging preferences and dynamically adjusts to better align with actual behaviors.
\item The EV Agent collects and analyzes technical parameters and dynamic data of EVs, such as battery capacity, charging duration, energy consumption, and SOC (state of charge) at the start and end of charging. By analyzing this data, the agent constructs a profile of users' vehicle and charging habits, providing decision support for optimal charging timing and methods while ensuring efficiency and cost-effectiveness.
\item The Psychology Agent builds on the user and vehicle profiles created by the previous agents, analyzing and constructing psychological traits such as time and price sensitivity. By incorporating users' economic status and historical charging data, the agent predicts the maximum acceptable delay time (\(t_{\text{tolerance}} \)) and provides insights into users' preferences for charging timing and methods, supporting personalized charging decisions.
\end{itemize}

\begin{figure}[htbp]
\includegraphics[width=\columnwidth]{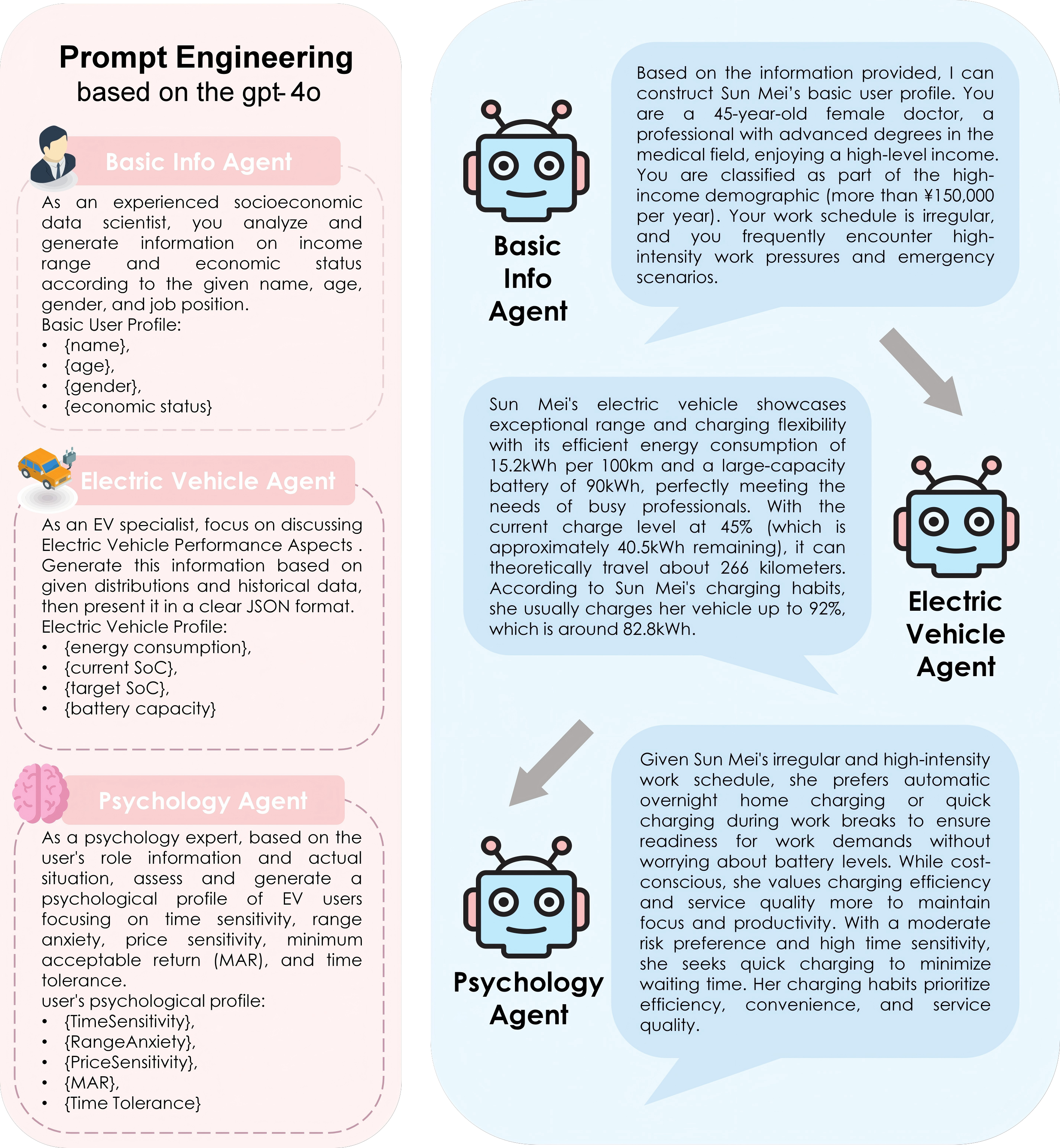}
\caption{User profile generation with large language model-based multi-agent systems.}
\label{fig:User profile generation with large language model-based multi-agent systems.}
\end{figure}

\subsection{User Behavioral Decision Agent}
In the previous stage, the agent generated a complete user profile, constructing a digital twin of the EV user. Building upon this, a user behavior decision model was developed. This agent interacts with the EVCS and DR models to make real-time decisions on whether the EV should participate in DR events and selects the most suitable charging strategy for the user. The key decision factors assessed by the agent include:

\begin{itemize}
\item The agent first checks whether the battery level is sufficient to meet the user's travel needs. If the current charge is insufficient, the model will suggest that the user participate in a DR event to compensate for the shortage in energy.
\item Based on the user profile, the agent evaluates whether the incentive amount offered by the EVCS is sufficiently attractive. If the incentive is compelling enough, the agent advocates for user participation in the DR event, even if it may entail a delay in charging.
\item The agent integrates user preferences and the availability of charging periods to recommend the most suitable charging strategy. For instance, tailored to individual users' charging requirements and time sensitivity, the model adjusts the charging strategy to optimize user experience.
\item The agent also evaluates the user's willingness to defer charging, particularly when there is a temporal misalignment between the current time and the periods of DR incentives. 
\end{itemize}

\begin{figure}[htbp]
\includegraphics[width=\columnwidth]{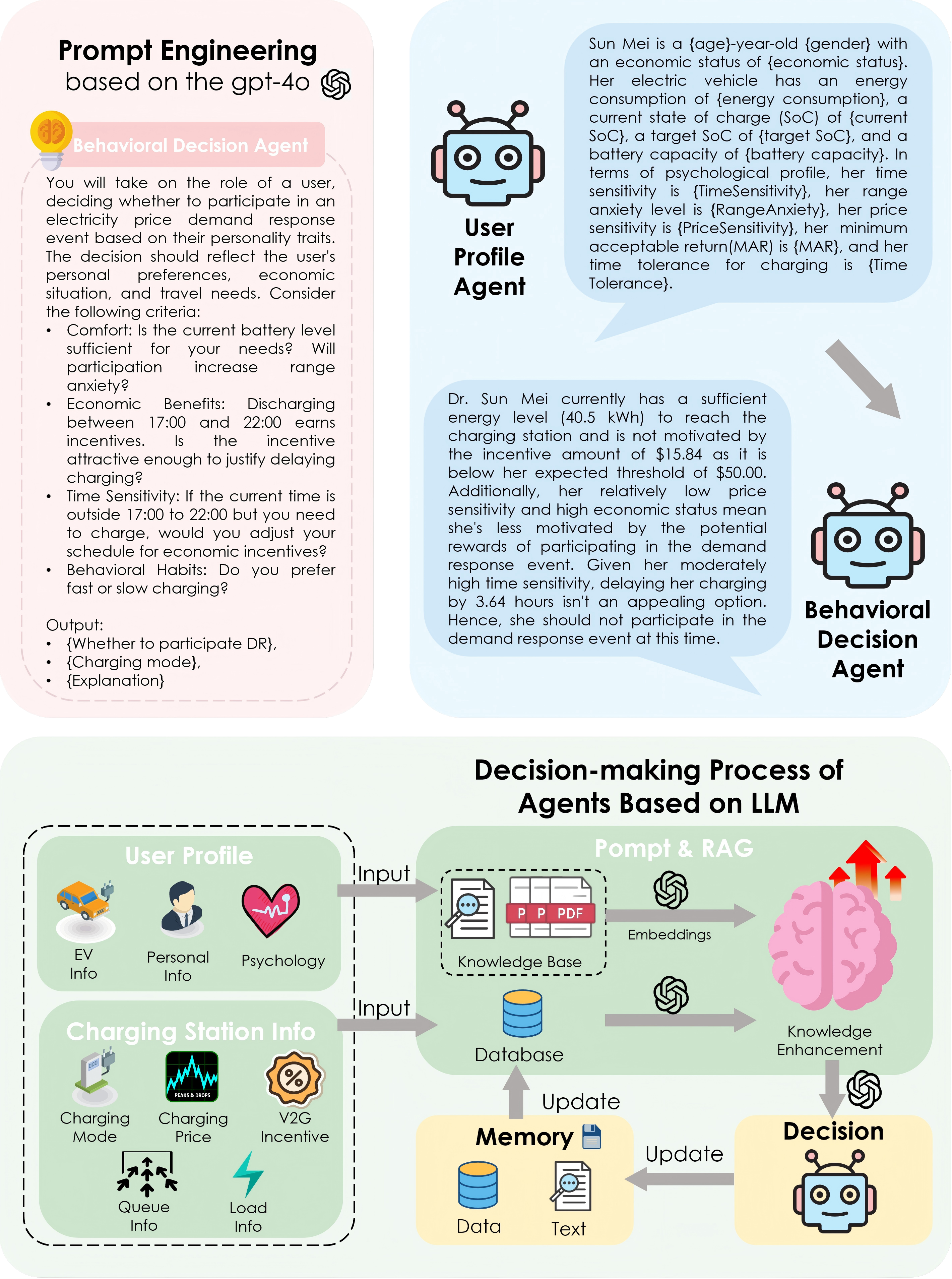}
\caption{User behavioral decision with large language model-based multi-agent systems.}
\label{fig:User behavioral decision with large language model-based multi-agent systems.}
\end{figure}

In the decision-making process, a minimum reward threshold is introduced to facilitate decision-making processes for the agent. To calculate the minimum reward a user is willing to accept for discharging (i.e., the minimum acceptable reward threshold), several key factors are considered, including battery degradation costs, inconvenience or loss due to delayed charging (i.e., time costs), opportunity costs, and the user's annual income level.

Considering potential delays in the charging and discharging process, the delay cost function \( C_{\text{delay}} \) is designed to quantify the inconvenience or actual loss caused by delayed charging. Assuming the user has a certain tolerance for delay \( t_{\text{tolerance}} \), the delay cost function is defined as follows:

{\footnotesize
\begin{equation}
\begin{aligned}
 C_{\text{delay}}(t) = \begin{cases} 
0 & \text{if } t \leq t_{\text{tolerance}} \\
a(t - t_{\text{tolerance}})^b & \text{otherwise}
\end{cases}
\end{aligned}\end{equation}}Where \( t \) represents the actual delay duration, \(t_{\text{tolerance}} \) is the maximum acceptable delay time, and \( a \) and \( b \) are constants that describe the rate of growth of the delay cost.

The minimum reward a user is willing to accept for discharging must cover all direct and indirect costs. Therefore, the minimum reward \( r_{\text{min}} \) is calculated as:

{\footnotesize
\begin{equation}
\begin{aligned}
r_{\text{min}} = \left( C_{\text{loss}} + C_{\text{delay}}(t) \right) \times \left( 1 + \gamma \right)
\end{aligned}\end{equation}}Where \( C_{\text{loss}} \) is the battery degradation cost, \( C_{\text{delay}}(t) \) is the delay cost, and \( \gamma \) is a risk premium factor that reflects the user's attitude towards uncertainty. The value of \( \gamma \) typically ranges from 0 to 1, based on the user's psychological traits (e.g., time sensitivity, price sensitivity). A more conservative user may opt for a higher \( \gamma \), thereby increasing the minimum reward requirement.

By integrating and analyzing a multitude of factors, the multi-agent model based on LLMs can dynamically adjust the charging strategies for EV s to respond to continuously changing external conditions. This approach more accurately emulates user charging behavior through personalized and detailed modeling. Such sophisticated modeling provides robust support for DR studies and related policy-making.

\subsection{Expert Evaluation Agents System}
We present a LLM-based multi-agent architecture for decision validation, incorporating specialized expert agents in economics, temporal management, power systems, consumer behavior, and comprehensive assessment. The system employs a Retrieval-Augmented Generation (RAG) framework integrating domain-specific knowledge graphs with real-time data streams to dynamically enhance agents' decision-making capabilities. Each agent operates autonomously while generating interpretable evaluation reports with traceable reasoning paths. A domain-expertise-weighted majority voting mechanism validates the rationality of proposed decisions through collective decision-making: approved decisions are directly outputted, while rejected ones trigger a feedback loop to refine the original decision through consensus-driven deliberation among all expert agents.
\section{\small{Simulation results}}
The system utilizes GPT-4 for question-and-answer dialogues to facilitate the decision-making process of user agents. Concurrently, a simulation environment for EVCS is developed using Python, enabling these agents to conduct decision-making simulations within this context. The primary objective is to investigate the feasibility of DR mechanisms.

\begin{table}[htbp]
\centering
\caption{Model Paramters}
\label{tab:configurations}
\begin{tabular}{clr}
\hline\hline
\multicolumn{1}{c}{\text{Variables}} & \multicolumn{1}{c}{\text{Definition}} & \multicolumn{1}{c}{\text{Value}} \\ \hline
$[T_0, T_1]$           & \makebox[3cm][c]{DR time window}                & [17:00, 22:00]           \\
$n$                 & \makebox[3cm][c]{
Number of EV users}                & 100         \\
$\Delta P_0^c$               & \makebox[3cm][c]{The nominal charging power}                        & 22kW        \\
$\Delta P_0^d$                 & \makebox[3cm][c]{The nominal discharging power}                                 & 15kW           \\ 
\hline\hline
\end{tabular}
\end{table}

In this study, the simulation results are as follow. Fig.5 reveals that at an incentive price of 0.2 RMB/kWh, there is already observable interest from users in participating in V2G programs. The willingness of users to participate increases with the rise in incentive prices. Notably, a significant surge in user engagement is observed when the incentive reaches 0.9 RMB/kWh. This phenomenon underscores that meeting users' expectations for returns through incentives can significantly enhance their inclination towards V2G project participation.

\begin{figure}[htbp]
\includegraphics[width=\columnwidth]{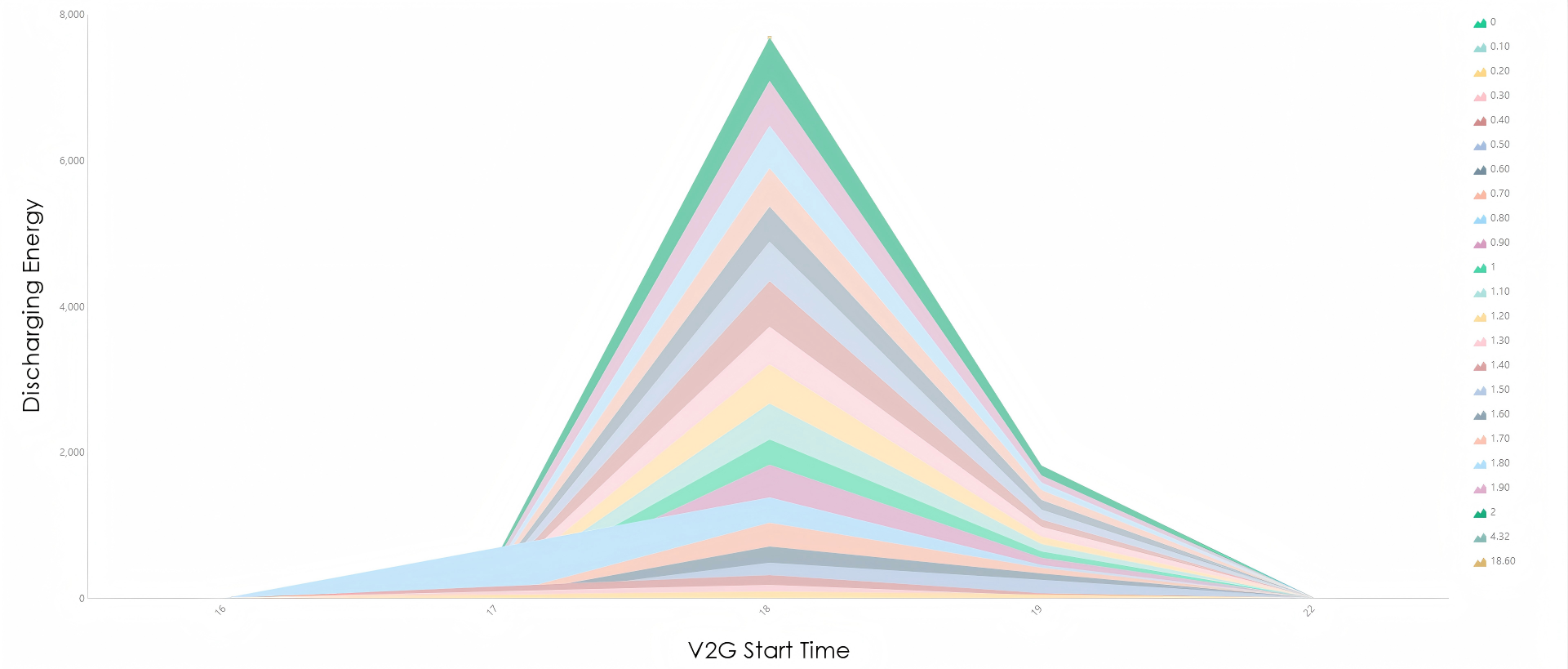}
\caption{The variation of electric vehicle  discharged energy under different incentive amounts across various time periods.}
\label{fig:The variation of electric vehicle  discharged energy under different incentive amounts across various time periods.}
\end{figure}
Further analysis of the data presented in Fig.6 indicates that, compared to the lack of coordination in EV charging behavior and subsequent peak power demand without any incentives, setting the incentive price within the range of 0.4 to 0.6 RMB/kWh leads to a marked reduction in peak power demand. Moreover, the volatility of power load tends to stabilize under these conditions. These findings suggest that the proposed DR mechanism not only effectively encourages EV users to actively engage in V2G projects but also facilitates orderly charging behaviors among users, thereby optimizing grid load management. However, it is crucial to note that excessively increasing the incentive price beyond this optimal range could result in delayed charging times due to overwhelming participation in V2G, subsequently causing a shift and increase in the amplitude of peak power demand. Therefore, appropriately setting the incentive price plays a critical role in the sustainable development of V2G initiatives.

\begin{figure}[htbp]
\includegraphics[width=\columnwidth]{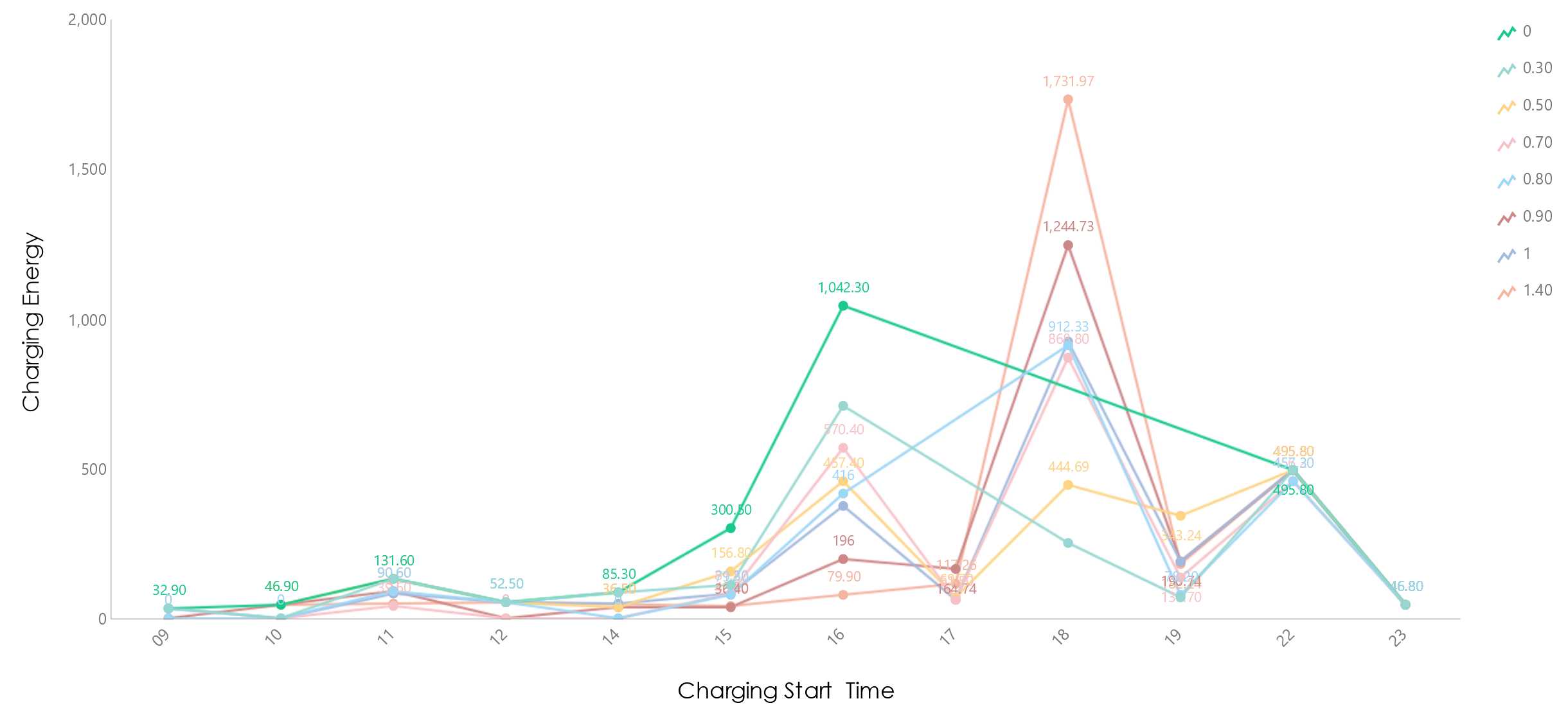}
\caption{The variation of electric vehicle charged energy under different incentive amounts across various time periods.}
\label{fig:The variation of electric vehicle charged energy under different incentive amounts across various time periods.}
\end{figure}

Research findings illustrated in Fig.7 show that the proposed DR mechanism effectively balances power supply and demand at EVCS, achieving peak load shaving and valley filling effects.Specifically, when the incentive electricity price is set at approximately 0.5 RMB/kWh, the peak load reduction effect of this DR strategy on EVCS loads reaches its optimum, efficiently mitigating load fluctuations. However, as the incentive price is further increased (e.g., reaching 0.9 RMB/kWh or higher), despite an increase in the number of responsive users, the cumulative effect caused by delayed charging behaviors paradoxically leads to a new round of substantial fluctuations in the total load of EVCS, creating new peak loads.

\begin{figure}[htbp]
\includegraphics[width=\columnwidth]{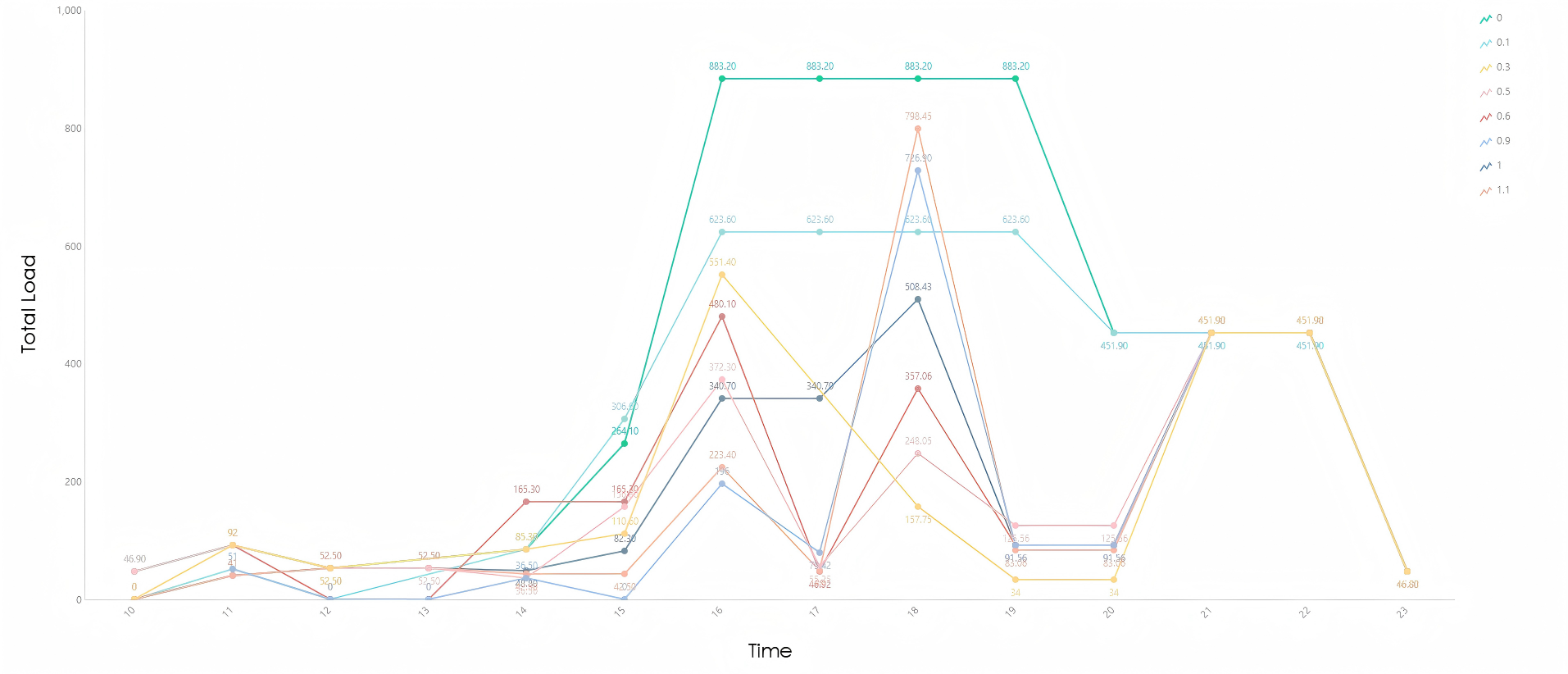}
\caption{The variation of electric vehicle charging station's total load under different incentive amounts across various time periods.}
\label{fig:The variation of electric vehicle charging station's total load under different incentive amounts across various time periods.}
\end{figure}

As observed from Fig.8, provided that the incentive price is controlled within 1.3 RMB/kWh, the overall revenue of EVCS can still be ensured. Therefore, when designing DR mechanisms to optimize the operational efficiency and economic benefits of EVCS, it is essential to consider the balance between the level of incentives and user response behaviors. This ensures not only the effective mobilization of user enthusiasm but also prevents new issues arising from over-incentivization.

\begin{figure}[htbp]
\includegraphics[width=\columnwidth]{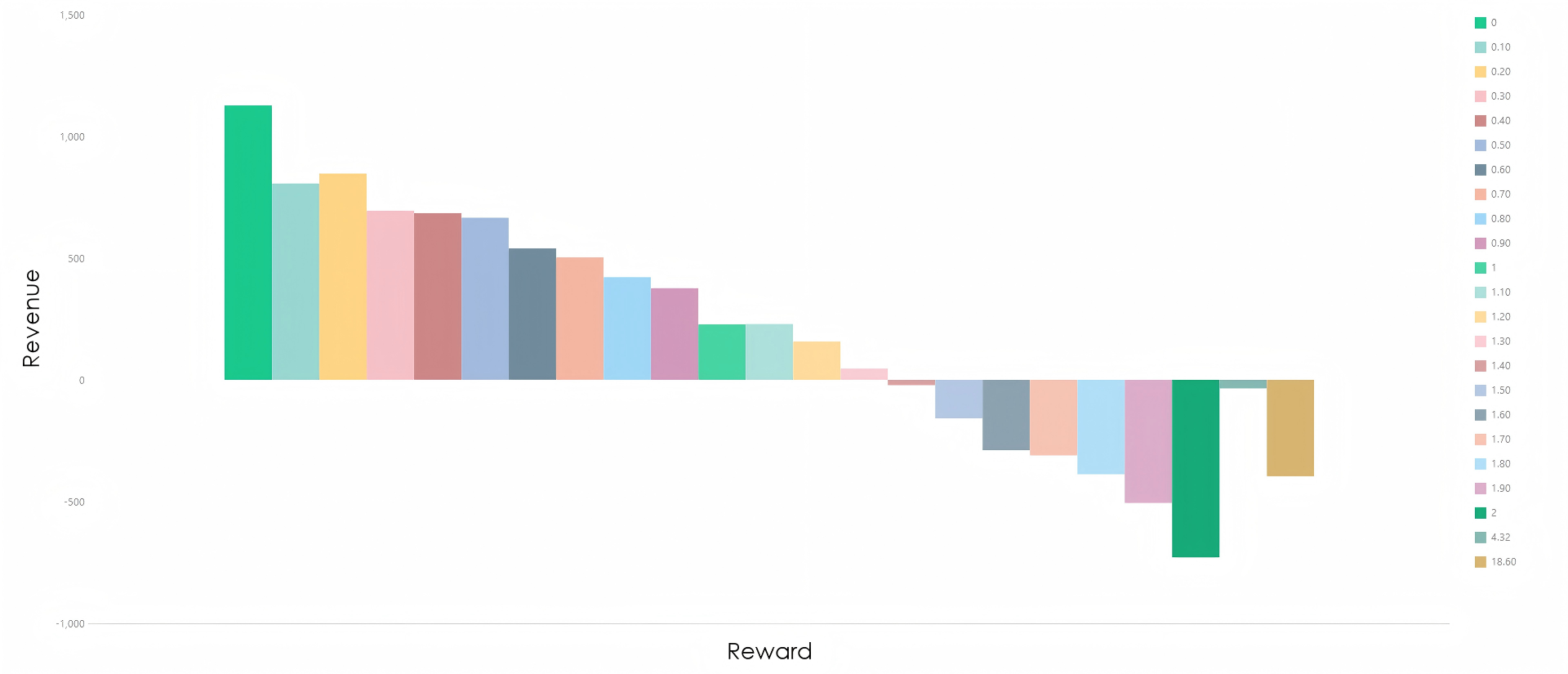}
\caption{Total revenue of the electric vehicle charging station under different incentive amounts.}
\label{fig:Total revenue of the electric vehicle charging station under different incentive amounts.}
\end{figure}
This approach emphasizes the necessity for a nuanced understanding of how financial incentives can shape consumer behavior while maintaining the operational integrity and profitability of charging infrastructures, critical insights for government policymakers and EVCS operators aiming to enhance the sustainability of EV  ecosystems.

\section{CONCLUSION}
This study introduces an innovative DR system enhanced by LLM-powered intelligent agents, offering personalized charging and discharging recommendations alongside dynamic pricing incentives. The proposed approach not only enhances grid stability and efficiency but also significantly boosts user satisfaction. Findings indicate that setting incentive prices within an optimal range  markedly increases user engagement in V2G initiatives, facilitating orderly charging behaviors and effectively mitigating peak load demands. However, careful calibration is essential to prevent the adverse effect of new peak loads caused by excessive participation. This LLM-driven DR framework provides a practical and sustainable strategy for policymakers and grid operators, contributing to the advancement of EV ecosystems.

\vspace{12pt}
\color{red}
IEEE conference templates contain guidance text for composing and formatting conference papers. Please ensure that all template text is removed from your conference paper prior to submission to the conference. Failure to remove the template text from your paper may result in your paper not being published.

\end{document}